# MAGNETO-OPTICAL PROPERTIES OF THE QUANTUM DOT – IMPURITY CENTER SYSTEMS SYNTHESIZED IN A TRANSPARENT DIELECTRIC MATRIX

**V.D. Krevchik[1], A.B. Grunin[1], A.K. Aringazin[2,3], M.B. Semenov[1,3]**

[1]**Department of Physics, Penza State University, Penza 440017, Russia**
physics@diamond.stup.ac.ru

[2]**Department of Theoretical Physics, Institute for Basic Research, Eurasian National University, Astana 473021, Kazakhstan**
aringazin@mail.kz

[3]**The Institute for Basic Research, P.O. Box 1577, Palm Harbor, FL 34682, USA**
ibr@gte.net

**Abstract**

Magneto-optical absorption by the quantum dot (QD) with impurity center (IC or $D^{(-)}$-center) complexes synthesized in a transparent dielectric matrix, with consideration of the QD size dispersion, is theoretically studied. Within the framework of effective mass approach, the analytical expression for the impurity magneto-optical absorption coefficient, for both the longitudinal and transversal polarizations with respect to the direction of external quantizing magnetic field, is calculated.

# 1 Introduction

Electro-optical [1-4] and magneto-optical [5-7] properties for quasi-zero-dimensional (quasi-0D) structures formed by the semiconductive sphere-shaped nanocrystals of radii about 1 to 100 nm synthesized in a transparent dielectric matrix are currently extensively studied. The interest is due to their similarity to heterogeneous phase systems which are new perspective materials that can be used to design nonlinear opto-electronic active elements, e.g., controlled (by optical signals) elements in quantum computers and lasers.

Magneto-optics of quantum dots (QD) attracts much attention because of the emerging new effects. These effects are related to a hybridization of dimensional and magnetic quantization [8]. On other hand, the existence of impurity centers in quasi-0D-structures stimulates an interest to the problem of controlled modulation of the binding energy of impurity centers [9] and, correspondingly, to the problem of heterogeneous phase systems with controlled QD impurity magneto-optical absorption.

In the present paper, we sketch results of theoretical study of the light impurity magneto-optical absorption for both the longitudinal and transversal polarizations (with respect to direction of an external magnetic field) in QD synthesized in a transparent dielectric matrix, with an account for dispersion of QD size [10-13]. Particularly, we investigate energy spectra of QD-$D^{(-)}$-center complex in a quantizing external magnetic field.

# 2 Features of the energy spectrum of QD-$D^{(-)}$-center complex in quantizing magnetic field

It is known [14] that $D^{(-)}$-states are a solid state analogous to that of $H^-$ ion. Such states in QD are of much interest since they are related to correlation effects in low-dimensional systems [14]. The $D^{(-)}$-centers-0D structure appears to exhibit a considerable increase of the binding energy [15] as compared with the three-dimensional (3D) case. In this section, we will show that one expects an enhance of population of the $D^{(-)}$-states in QD because of the hybrid quantization. We will consider the semiconductive sphere-shaped QD with the radius $R_0$ in quantizing external magnetic field. Calculations will be made in the cylindrical system of reference, with the origin $O$ in QD-center and the magnetic induction vector $\vec{B}$ being pointed along the $z$-axis, $\vec{B} \uparrow\uparrow \vec{k}$, where $\vec{k}$ is the unit vector along the $z$-axis. To describe one-electron states in QD we use the spherical oscillator well confinement potential of the following form:

$$V_0(\rho, z) = \frac{m^* \omega_0^2 \left(\rho^2 + z^2\right)}{2}, \qquad (1)$$

where $m^*$ is the electron effective mass, $\omega_0$ is the QD binding potential characteristic frequency, and $\rho, \varphi, z$ are cylindrical coordinates; $\rho \le R_0$, $-R_0 \le z \le R_0$.

Within the effective mass approximation, in the symmetrical gauge fixing of the vector-potential $\vec{A}$ the Hamiltonian operator $H_{QD}$ in the chosen system of reference can be written as

$$H_{QD} = -\frac{\hbar^2}{2m^*}\left(\frac{1}{\rho}\frac{\partial}{\partial\rho}\left(\rho\frac{\partial}{\partial\rho}\right) + \frac{1}{\rho^2}\frac{\partial^2}{\partial\varphi^2}\right) - \frac{i\hbar\omega_B}{2}\frac{\partial}{\partial\varphi} + \frac{m^*}{2}\left(\omega_0^2 + \frac{\omega_B^2}{4}\right)\rho^2 + H_{zQD}, \qquad (2)$$

where $\omega_B = |e| B / m^*$ is the cyclotron frequency, $|e|$ is the charge of electron; $B$ is the absolute value of the magnetic induction $\vec{B}$, and

$$H_{zQD} = -\hbar^2 / \left(2m^*\right)\left(\partial^2 / \partial z^2\right) + m^* \omega_0^2 z^2 / 2 .$$

Eigenvalues $E_{n_1, m, n_2}$ and corresponding eigenfunctions $\Psi_{n_1, m, n_2}(\rho, \varphi, z)$ of the Hamiltonian (2) are given by the following expressions [16]:

$$E_{n_1,m,n_2}=\frac{\hbar\omega_B m}{2}+\hbar\omega_0\left(n_2+\frac{1}{2}\right)+\hbar\omega_0\sqrt{1+\frac{\omega_B^2}{4\omega_0^2}}\left(2n_1+|m|+1\right), \quad (3)$$

$$\Psi_{n_1,m,n_2}(\rho,\varphi,z)=\frac{1}{a_1}\left(\frac{n_1!}{2^{n_2+1}\pi^{\frac{3}{2}}n_2!\left(n_1+|m|\right)!a}\right)^{\frac{1}{2}}\left(\frac{\rho^2}{2a_1^2}\right)^{\frac{|m|}{2}}\exp\left[-\left(\frac{\rho^2}{4a_1^2}+\frac{z^2}{2a^2}\right)\right]\times$$

$$\times H_{n_2}\left(\frac{z}{a}\right)L_{n_1}^{|m|}\left(\frac{\rho^2}{2a_1^2}\right)\exp(im\varphi). \quad (4)$$

Here, $n_1, n_2 = 0,1,2,...$ are quantum numbers corresponding to Landau levels and to energy levels for spherically-symmetric oscillator potential well, respectively, $m = 0, \pm 1, \pm 2,...$ is the magnetic quantum number, $a_1^2 = a^2/\left(2\sqrt{1+a^4/\left(4a_B^4\right)}\right)$, $a=\sqrt{\hbar/\left(m^*\omega_0\right)}$, $a_B=\sqrt{\hbar/\left(m^*\omega_B\right)}$ is the magnetic length, $H_{n_2}(x)$ is Hermite polynomial [17].

Let us consider $D^{(-)}$-center localized at the point $\vec{R}_a=(\rho_a,\varphi_a,z_a)$. The impurity potential is modeled by the zero-range potential with the intensity $\gamma = 2\pi\hbar^2/\left(\alpha m^*\right)$. In the cylindrical system of reference, this potential can be written as

$$V_\delta(\rho,\varphi,z;\rho_a,\varphi_a,z_a)=\gamma\frac{\delta(\rho-\rho_a)}{\rho}\delta(\varphi-\varphi_a)\delta(z-z_a)\left[1+(\rho-\rho_a)\frac{\partial}{\partial\rho}+(z-z_a)\frac{\partial}{\partial z}\right], \quad (5)$$

where $\alpha$ is determined by the binding energy $E_i$ for the same $D^{(-)}$-center in massive semiconductor and $\delta(x)$ is Dirac delta-function. The wave-function $\Psi_{\lambda_B}^{(QD)}(\rho,\varphi,z;\rho_a,\varphi_a,z_a)$ for the electron, which is localized at the $D^{(-)}$-center, in the effective mass approximation, satisfies Schroedinger equation,

$$\left(E_{0\lambda_B}-H_{QD}\right)\Psi_{\lambda_B}^{(QD)}(\rho,\varphi,z;\rho_a,\varphi_a,z_a)=V_\delta(\rho,\varphi,z;\rho_a,\varphi_a,z_a)\Psi_{\lambda_B}^{(QD)}(\rho,\varphi,z;\rho_a,\varphi_a,z_a), \quad (6)$$

where $E_{0\lambda_B} = -\hbar^2\lambda_B^2/(2m^*)$ are eigenvalues of the Hamiltonian

$$H^{\delta}_{(QD)B} = H_{QD} + V_\delta(\rho,\varphi,z;\rho_a,\varphi_a,z_a).$$

One-electron Green function $G(\rho,\varphi,z,\rho_1,\varphi_1,z_1;E_{0\lambda_B})$ associated to the Schroedinger equation (6), corresponding to the energy $E_{0\lambda_B}$ and to the source at point $\vec{r}_1 = (\rho_1,\varphi_1,z_1)$, can be written as

$$G(\rho,\varphi,z,\rho_1,\varphi_1,z_1;E_{0\lambda_B}) = \sum_{n_1,m,n_2} \frac{\Psi^*_{n_1,m,n_2}(\rho_1,\varphi_1,z_1)\Psi_{n_1,m,n_2}(\rho,\varphi,z)}{(E_{0\lambda_B} - E_{n_1,m,n_2})}. \quad (7)$$

The Lippman-Schwinger equation for $D^{(-)}$-state in QD in the external magnetic field is

$$\Psi^{(QD)}_{\lambda_B}(\rho,\varphi,z;\rho_a,\varphi_a,z_a) = \int_{-\infty}^{+\infty}\int_0^{2\pi}\int_0^{+\infty}\rho_1\, d\rho_1\, d\varphi_1\, dz_1\, G(\rho,\varphi,z,\rho_1,\varphi_1,z_1;E_{0\lambda_B}) \times$$
$$\times V_\delta(\rho_1,\varphi_1,z_1;\rho_a,\varphi_a,z_a)\, \Psi^{(QD)}_{\lambda_B}(\rho_1,\varphi_1,z_1;\rho_a,\varphi_a,z_a). \quad (8)$$

Substituting the zero-range potential (5) into Eq. (8) we obtain

$$\Psi^{(QD)}_{\lambda_B}(\rho,\varphi,z;\rho_a,\varphi_a,z_a) = \gamma G(\rho,\varphi,z,\rho_a,\varphi_a,z_a;E_{0\lambda_B}) \times$$
$$\times\left(\hat{T}\Psi^{(QD)}_{\lambda_B}\right)(\rho_a,\varphi_a,z_a;\rho_a,\varphi_a,z_a), \quad (9)$$

where

$$\left(\hat{T}\Psi^{(QD)}_{\lambda_B}\right)(\rho_a,\varphi_a,z_a;\rho_a,\varphi_a,z_a) \equiv$$
$$\equiv \lim_{\substack{\rho\to\rho_a\\ \varphi\to\varphi_a\\ z\to z_a}} \left[1 + (\rho-\rho_a)\frac{\partial}{\partial\rho} + (z-z_a)\frac{\partial}{\partial z}\right]\Psi^{(QD)}_{\lambda_B}(\rho,\varphi,z;\rho_a,\varphi_a,z_a). \quad (10)$$

The equation, which determines dependence of the energy $E_{0\lambda_B}$ of the bound state (for $D^{(-)}$-center) on the parameters of QD, position $\vec{R}_a = (\rho_a,\varphi_a,z_a)$ of the impurity, and magnetic field $B$ can be obtain by applying the operator $T$ to the left and right hand sides of Eq. (9). This gives us

$$\alpha=\frac{2\pi\hbar^2}{m^*}\left(\hat{T}G\right)\left(\rho_a,\varphi_a,z_a,\rho_a,\varphi_a,z_a;E_{0\lambda_B}\right), \tag{11}$$

Let us consider the case when the impurity level $E_{0\lambda_B}$ is lower than the bottom of the QD potential well, i.e., $E_{0\lambda_B}<0$. Then, in the units of effective Bohr energy $E_d$ and the effective Bohr radius $a_d=4\pi\varepsilon_0\varepsilon\hbar^2/\left(m^*|e|^2\right)$, where $\varepsilon$ is the QD relative static dielectric permeability, the Green function $G\left(\rho,\varphi,z,\rho_a,\varphi_a,z_a;E_{0\lambda_B}\right)$ can be written as

$$G\left(\rho,\varphi,z,\rho_a,\varphi_a,z_a;E_{0\lambda_B}\right)=-\frac{\beta_1}{2\pi^{\frac{3}{2}}aa_1^2E_d}\exp\left[-\left(\frac{\rho_a^2+\rho^2}{4a_1^2}+\frac{z_a^2+z^2}{2a^2}\right)\right]\times$$

$$\times\int_0^{+\infty}dt\,\exp\left[-\left(\beta_1\eta_{0B}^2+w_1+\frac{1}{2}\right)t\right]\sum_{n_2=0}^{\infty}\left(\frac{e^{-t}}{2}\right)^{n_2}\frac{H_{n_2}\left(\frac{z_a}{a}\right)H_{n_2}\left(\frac{z}{a}\right)}{n_2!}\times$$

$$\times\sum_{m=-\infty}^{+\infty}\exp\left[-|m|w_1t\right]\exp\left[\left(i(\varphi-\varphi_a)-\beta_1a^{*-2}t\right)m\right]\left(\frac{\rho_a\rho}{2a_1^2}\right)^{|m|}\times$$

$$\times\sum_{n_1=0}^{\infty}\frac{n_1!}{\left(n_1+|m|\right)!}L_{n_1}^{|m|}\left(\frac{\rho_a^2}{2a_1^2}\right)L_{n_1}^{|m|}\left(\frac{\rho^2}{2a_1^2}\right)\exp\left[-2n_1w_1t\right], \tag{12}$$

where $\beta_1=R_0^*/\left(4\sqrt{U_0^*}\right)$, $R_0^*=2R_0/a_d$, $U_0^*=U_0/E_d$, $\eta_{0B}^2=\left|E_{0\lambda_B}\right|/E_d$, $w_1=\sqrt{1+\beta_1^2a^{*-4}}$, and $a^*=a_B/a_d$.

Summation over quantum number $n_2$ in Eq. (12) can be performed using Mehler formula [18],

$$\sum_{n_2=0}^{\infty}\left(\frac{e^{-t}}{2}\right)^{n_2}\frac{H_{n_2}\left(\frac{z_a}{a}\right)H_{n_2}\left(\frac{z}{a}\right)}{n_2!}=\frac{1}{\sqrt{1-e^{-2t}}}\exp\left\{\frac{2z_a z e^{-t}-\left(z_a^2+z^2\right)e^{-2t}}{a^2\left(1-e^{-2t}\right)}\right\}.$$

(13)

Also, Hille-Hardi formula for bilinear generating function [18] can be used for the summation over quantum number $n_1$,

$$\sum_{n_1=0}^{\infty}\frac{n_1!}{\left(n_1+|m|\right)!}L_{n_1}^{|m|}\left(\frac{\rho_a^2}{2a_1^2}\right)L_{n_1}^{|m|}\left(\frac{\rho^2}{2a_1^2}\right)\exp\left[-2n_1 w_1 t\right]=\left(\frac{\rho_a\rho}{2a_1^2}\right)^{-|m|}\exp\left[|m|w_1 t\right]\times$$
$$\times\left(1-\exp\left[-2w_1 t\right]\right)^{-1}\exp\left[-\exp\left[-2w_1 t\right]\frac{\left(\rho_a^2+\rho^2\right)}{2a_1^2\left(1-\exp\left[-2w_1 t\right]\right)}\right]\times$$
$$\times I_{|m|}\left(\frac{\rho_a\rho\exp\left[-w_1 t\right]}{a_1^2\left(1-\exp\left[-2w_1 t\right]\right)}\right). \tag{14}$$

The result of summation over magnetic quantum number $m$ can be written as

$$\sum_{m=-\infty}^{+\infty}\exp\left[\left(i\left(\varphi-\varphi_a\right)-\beta_1 a^{*-2}t\right)m\right]I_{|m|}\left(\frac{\rho_a\rho\exp\left[-w_1 t\right]}{a_1^2\left(1-\exp\left[-2w_1 t\right]\right)}\right)=$$
$$=\exp\left[\frac{1}{2}\left(\exp\left[i(\varphi-\varphi_a)-\beta_1 a^{*-2}t\right]+\exp\left[-i(\varphi-\varphi_a)+\beta_1 a^{*-2}t\right]\right)\frac{\rho_a\rho\exp\left[-w_1 t\right]}{a_1^2\left(1-\exp\left[-2w_1 t\right]\right)}\right].$$

(15)

Thus, using Eqs. (13) – (15) we obtain the Green function (12) in the following cumbersome form:

$$G\left(\rho,\varphi,z,\rho_a,\varphi_a,z_a;E_{0\lambda_B}\right)=-\frac{1}{2^3\pi^{\frac{3}{2}}\sqrt{\beta_1}E_d a_d^3}\left[\int_0^{+\infty}dt\exp\left[-\left(\beta_1\eta_{0B}^2+w_1+\frac{1}{2}\right)t\right]\times\right.$$

$$\times\left(2\sqrt{2}\,w_1\exp\left[-\frac{z_a^2+z^2}{4\beta_1 a_d^2}\right]\left(1-e^{-2t}\right)^{-\frac{1}{2}}\left(1-\exp\left[-2w_1 t\right]\right)^{-1}\times\right.$$

$$\times\exp\left\{\frac{2z_a z e^{-t}-\left(z_a^2+z^2\right)e^{-2t}}{2\beta_1 a_d^2\left(1-e^{-2t}\right)}\right\}\exp\left[-\frac{\left(\rho_a^2+\rho^2\right)w_1\left(1+\exp\left[-2w_1 t\right]\right)}{4\beta_1 a_d^2\left(1-\exp\left[-2w_1 t\right]\right)}\right]\times$$

$$\times\exp\left[\frac{1}{2}\left(\exp\left[i(\varphi-\varphi_a)-\beta_1 a^{*-2}t\right]+\exp\left[-i(\varphi-\varphi_a)+\beta_1 a^{*-2}t\right]\right)\frac{\rho_a\rho w_1\exp\left[-w_1 t\right]}{\beta_1 a_d^2\left(1-\exp\left[-2w_1 t\right]\right)}\right]-$$

$$\left.-t^{-\frac{3}{2}}\exp\left[-\frac{\left(\rho-\rho_a\right)^2 w_1+\left(z-z_a\right)^2}{4\beta_1 a_d^2 t}\right]\right)+$$

$$\left.+2\sqrt{\pi\beta_1}\,a_d\,\frac{\exp\left[-\sqrt{\frac{\left(2\beta_1\eta_{0B}^2+2w_1+1\right)\left(\left(\rho-\rho_a\right)^2 w_1+\left(z-z_a\right)^2\right)}{2\beta_1 a_d^2}}\right]}{\sqrt{\left(\rho-\rho_a\right)^2 w_1+\left(z-z_a\right)^2}}\right]. \quad (16)$$

Inserting Eq. (16) into Eq. (11) gives

$$\sqrt{\eta_{0B}^2+\left(2\beta_1\right)^{-1}+w_1\beta_1^{-1}}=\eta_i-\sqrt{\frac{2}{\pi\beta_1}}\int_0^{+\infty}dt\exp\left[-\left(\beta_1\eta_{0B}^2+w_1+\frac{1}{2}\right)t\right]\times$$

$$\times\left[\frac{1}{2t\sqrt{2t}}-w_1\left(1-e^{-2t}\right)^{-\frac{1}{2}}\left(1-\exp\left[-2w_1 t\right]\right)^{-1}\exp\left[-\frac{z_a^{*2}\left(1-e^{-t}\right)}{2\beta_1\left(1+e^{-t}\right)}\right]\times\right.$$

$$\times\exp\left[-\frac{\rho_a^{*2}w_1}{2\beta_1\left(1-\exp\left[-2w_1 t\right]\right)}\left(1+\exp\left[-2w_1 t\right]-2\exp\left[-w_1 t\right]\times\right.\right.$$

$$\left.\left.\left.\times\left(\exp\left[-\beta_1 a^{*-2}t\right]+\exp\left[\beta_1 a^{*-2}t\right]\right)\right)\right]\right], \quad (17)$$

where $\eta_i^2=\left|E_i\right|/E_d$, $z_a^*=z_a/a_d$, and $\rho_a^*=\rho_a/a_d$. We observe a drastic modification of the QD electron states related to the dimensional quantization in three directions as it yields anisotropy of the binding energy of $D^{(-)}$-center. Namely, in the plane perpendicular to the magnetic field direction there is a dimensional quantization. For the $D^{(-)}$-centers, which are situated in the radial plane, the binding energy $\left(E_{\lambda_B}^{(QD)}\right)_\rho$ can be represented as

$$\left(E_{\lambda_B}^{(QD)}\right)_\rho = \begin{cases} \hbar\omega_0\sqrt{1+\dfrac{\omega_B^2}{4\omega_0^2}}+\left|E_{0\lambda_B}\right|, E_{0\lambda_B}<0, \\ \hbar\omega_0\sqrt{1+\dfrac{\omega_B^2}{4\omega_0^2}}-E_{0\lambda_B}, E_{0\lambda_B}>0 \end{cases} \tag{18}$$

Along the magnetic field direction, the $D^{(-)}$-center binding energy $\left(E_{\lambda_B}^{(QD)}\right)_z$ is found as

$$\left(E_{\lambda_B}^{(QD)}\right)_z = \begin{cases} \dfrac{\hbar\omega_0}{2}+\left|E_{0\lambda_B}\right|, E_{0\lambda_B}<0, \\ \dfrac{\hbar\omega_0}{2}-E_{0\lambda_B}, E_{0\lambda_B}>0 \end{cases} \tag{19}$$

As numerical analysis of the expression (18) shows, dependence of the binding energy $\left(E_{\lambda_B}^{(QD)}\right)_\rho$ (for the QD - $D^{(-)}$-center complex based on InSb) on the polar radius $\rho_a^* = \rho_a / a_d$ (in Bohr units) for $E_{0\lambda_B} < 0$ (Fig. 1) resembles the corresponding dependence for the case of QW-$D^{(-)}$-center system [22]. Fig. 2 present the dependence of the binding energy $\left(E_{\lambda_B}^{(QD)}\right)_z$ (for the QD-$D^{(-)}$-center complex based on InSb) on the coordinate $z_a^* = z_a / a_d$ (in Bohr units) for $E_{0\lambda_B} < 0$. As one can see from Fig. 2, in the case when impurity levels are lower than the bottom level of QD the $D^{(-)}$-center binding energy slightly decreases with an increase of magnetic field (compare curves 1 and 2 of Fig. 2). This is due to the absence of dependence of the ground oscillator state (along the *z*-axis) on the intensity of magnetic field, i.e., due to the absence of magnetic quantization. Hence, in this direction the magnetic field gives unstable influence on the QD $D^{(-)}$-states.

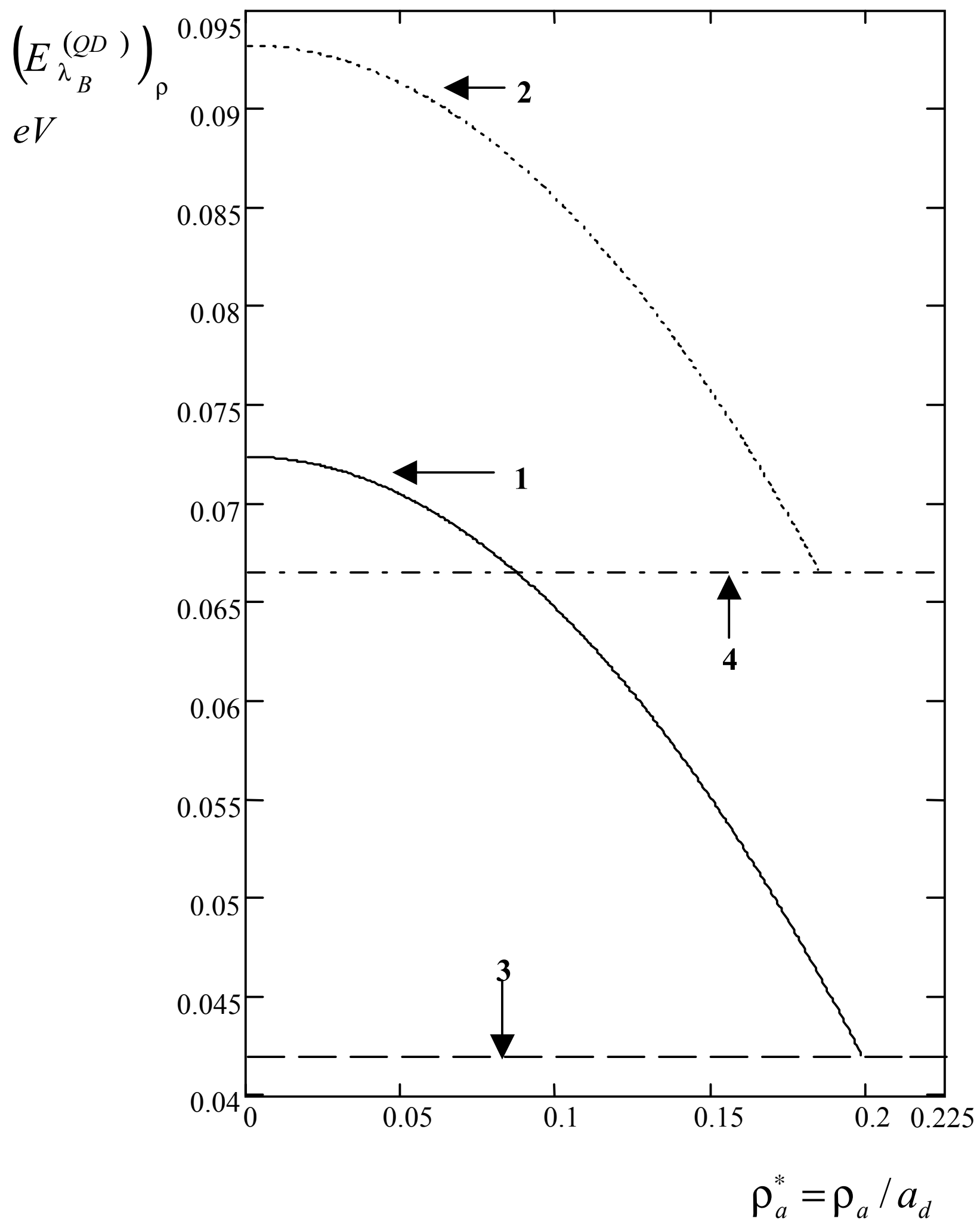

Fig. 1. The binding energy $\left(E_{\lambda_B}^{(QD)}\right)_\rho$ ($E_{0\lambda_B} < 0$) for the QD - D$^{(-)}$-center complex based on InSb as a function of the impurity polar radius $\rho_a^* = \rho_a / a_d$ for different values of magnetic field $B$. Curves 3 and 4 show the two-dimensional oscillator ground state energy levels positions for $B$=0T and $B$=12T, respectively; $|E_i| = 3.5\times10^{-2}$ $eV$, $R_0 = 35.8$ $nm$, $U_0 = 0.2$ $eV$): 1 – $B$=0 T; 2 – $B$=12 T .

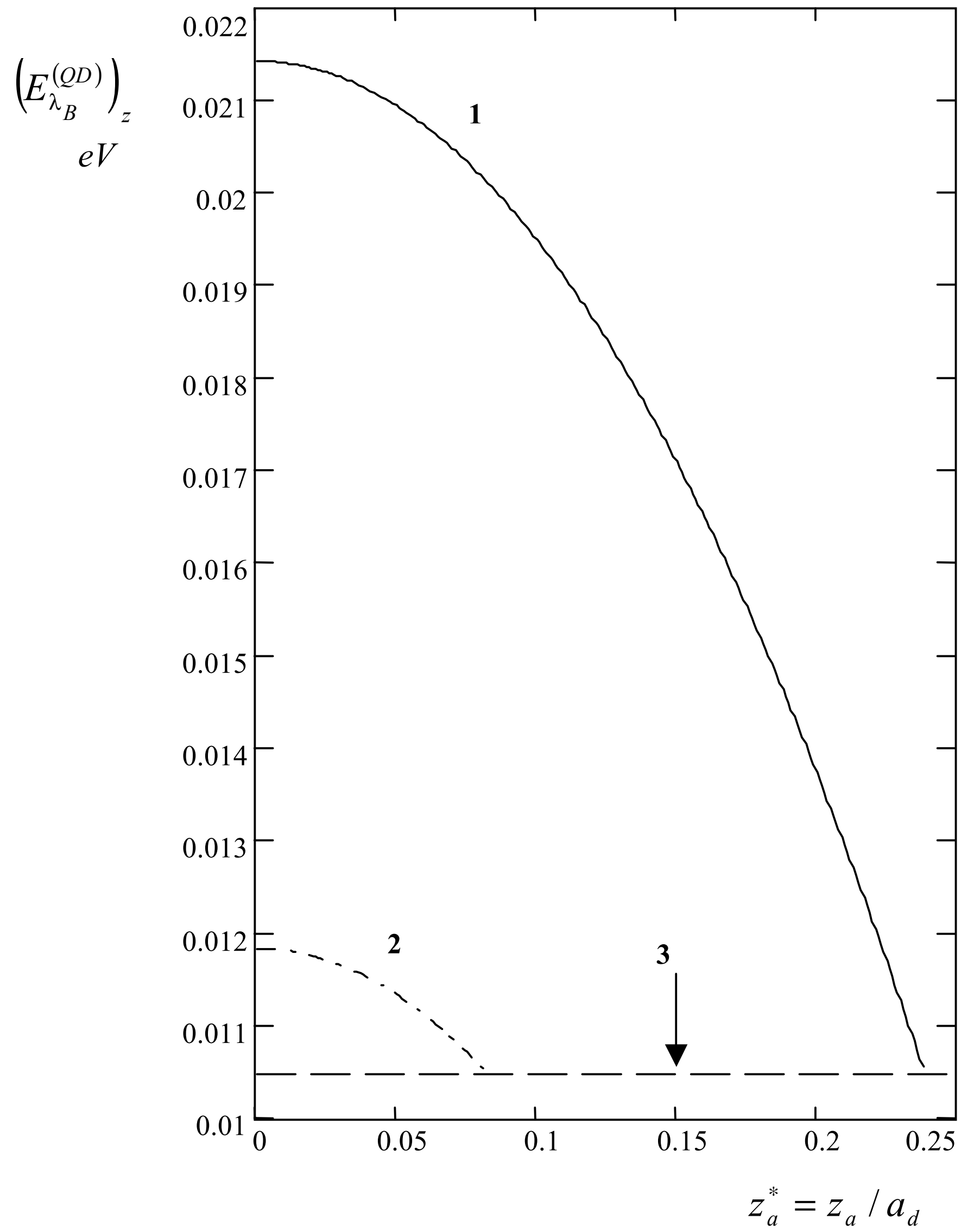

Fig. 2. The binding energy $\left(E_{\lambda_B}^{(QD)}\right)_z$ ($E_{0\lambda_B} < 0$) dependence, (for $D^{(-)}$-center in QD, based on InSb), from coordinate $z_a^* = z_a / a_d$, (line 3 shows the one-dimensional oscillator ground state energy level); $|E_i| = 1.38 \times 10^{-2}\ eV$, $R_0 = 71.6\ nm$, $U_0 = 0.2\ eV$). Curve 1: $B = 0\ T$, curve 2: $B = 15\ T$.

## 3 The effect of hybridization of the dimensional and magnetic quantization in light impurity absorption spectra

Magneto-optical properties of heterogeneous phase systems are of much interest due to the arising possibility to observe experimentally the effect of hybridization of the dimensional and magnetic quantizations in the impurity absorption spectra. In this section we will show that this effect provides important information on the zone structure and on the QD impurity states. This information can be obtained, for example, from analysis of the Zeeman energy shift and the period of oscillations in the magneto-optical impurity absorption spectra.

Let us consider the impurity absorption by the QD-$D^{(-)}$ center complex synthesized in a transparent dielectric matrix, in the case of ***longitudinal*** polarization with respect to the direction of the external magnetic field, $\vec{e}_\lambda \uparrow\uparrow \vec{B}$, where $\vec{e}_\lambda$ is the light polarization vector of unit length. Let us also consider the $D^{(-)}$-center, which is localized at the point $\vec{R}_a = (0,0,0)$. The impurity binding state energy level $E_{0\lambda_B}$ is assumed to be lower than the bottom of the spherical oscillator well. This well corresponds to the QD confinement potential, $E_{0\lambda_B} < 0$. The wave function $\Psi_{\lambda_B}^{(QD)}(\rho,\varphi,z;0)$ of the electron localized in the short-range potential of the $D^{(-)}$-center can be written as (see Eq. (16)):

$$\Psi_{\lambda_B}^{(QD)}(\rho,\varphi,z;0) = C_{1B}^{QD}\int_0^{+\infty} dt \exp\left[-\left(\beta_1\eta_{0B}^2 + w_1 + \frac{1}{2}\right)t\right]\left(1-e^{-2t}\right)^{-\frac{1}{2}}\left(1-\exp[-2w_1 t]\right)^{-1}\times$$
$$\times\exp\left[-\frac{z^2}{4\beta_1 a_d^2}\frac{\left(1+e^{-2t}\right)}{\left(1-e^{-2t}\right)}\right]\exp\left[-\frac{\rho^2 w_1}{4\beta_1 a_d^2}\frac{\left(1+\exp[-2w_1 t]\right)}{\left(1-\exp[-2w_1 t]\right)}\right]. \quad (20)$$

Here, $\Psi_{\lambda_B}^{(QD)}(\rho,\varphi,z;0) \equiv \Psi_{\lambda_B}^{(QD)}(\rho,\varphi,z;0,0,0)$ and the coefficient $C_{1B}^{QD} = \pi^{-\frac{3}{2}} w_1 C_B^{QD}$ is determined by

$$C_B^{QD}=\left[-2^{-\frac{1}{2}}\pi^{-\frac{3}{2}}\beta_1^{\frac{3}{2}}a_d^3 w_1\Gamma\left(\frac{1}{2}-w_1\right)\frac{\Gamma\left(\frac{\beta_1\eta_{0B}^2+w_1}{2}+\frac{5}{4}\right)}{\left(\frac{\beta_1\eta_{0B}^2+w_1}{2}+\frac{1}{4}\right)^2\Gamma\left(\frac{\beta_1\eta_{0B}^2-w_1}{2}+\frac{3}{4}\right)}\times\right.$$

$$\left.\times\left[\left(\frac{\beta_1\eta_{0B}^2+w_1}{2}+\frac{1}{4}\right)\left[\Psi\left(\frac{\beta_1\eta_{0B}^2+w_1}{2}+\frac{5}{4}\right)-\Psi\left(\frac{\beta_1\eta_{0B}^2-w_1}{2}+\frac{3}{4}\right)\right]-1\right]\right]^{-\frac{1}{2}}. \quad (21)$$

The effective Hamiltonian $H_{\text{int}B}^{(s)}$ of interaction with the light wave field in the case of longitudinal polarization $\vec{e}_{\lambda s}$, with respect to the direction of external magnetic field, can be written as

$$H_{\text{int}B}^{(s)}=-i\hbar\lambda_0\sqrt{\frac{2\pi\hbar^2\alpha^*}{m^{*2}\omega}I_0}\exp\left(i\vec{q}_s\vec{r}\right)\left(\vec{e}_{\lambda s}\nabla_{\vec{r}}\right), \quad (22)$$

where $\lambda_0$ is the local field coefficient, $\alpha^*$ is the fine structure constant with account for the static relative dielectric permeability $\varepsilon$, $I_0$ is the intensity of electromagnetic wave with frequency $\omega$, wave vector $\vec{q}_s$ and the polarization unit vector $\vec{e}_{\lambda s}$, and $\nabla_{\vec{r}}$ is the Hamiltonian operator.

Calculation of the matrix elements $M_{fQD,\lambda_B}^{(s)}$ for the dipole optical transitions in the case of the longitudinal light polarization $\vec{e}_{\lambda s}$ leads to integrals of the following types:

$$\int_0^{2\pi}\exp(-im\varphi)d\varphi=\begin{cases}0, if\ m\neq 0,\\ 2\pi, if\ m=0,\end{cases}$$

$$\int_{-\infty}^{+\infty}dz\,z\exp\left[-\frac{z^2}{2\beta_1 a_d^2\left(1-e^{-2t}\right)}\right]H_{n_2}\left(\frac{z}{\sqrt{2\beta_1}a_d}\right)=$$

$$=\begin{cases} 0, if\ n_2 \neq 2n+1, n=0,1,2,..., \\ (-1)^n\, 2^{2n+2}\beta_1\, a_d^2\Gamma\left(n+\frac{3}{2}\right)\exp[-2\,n\,t]\left(1-e^{-2t}\right)^{\frac{3}{2}}, if\ n_2 = 2n+1. \end{cases} \tag{23}$$

Due to Eq. (23) optical transitions from the $D^{(-)}$-center ground state occurs only to the QD states with $m$=0 and odd quantum number $n_2$ values. Taking into account for the pointed out selection rules, we write the matrix elements $M^{(s)}_{f\,QD,\lambda_B}$ of the considered optical transitions,

$$M^{(s)}_{f\,QD,\lambda_B} = \pi^{\frac{3}{2}} i\lambda_0\sqrt{\frac{\alpha^* I_0}{\omega}}E_d\, a_d^4\beta_1 w_1^{-1}\, C_{1B}^{QD}\, C_{n_1,0,2n+1}\left(-1\right)^n 2^{2n+\frac{5}{2}}\Gamma\left(n+\frac{3}{2}\right)\times$$
$$\times\frac{\left(2n+3/2+\left(2n_1+1\right)w_1+\beta_1\eta_{0B}^2\right)}{\left(\frac{\beta_1\eta_{0B}^2}{2}+n+\left(n_1+\frac{1}{2}\right)w_1+\frac{1}{4}\right)\left(\frac{\beta_1\eta_{0B}^2}{2}+n+\left(n_1+\frac{1}{2}\right)w_1+\frac{5}{4}\right)}, \tag{24}$$

where the normalization constant is

$$C_{1B}^{QD}C_{n_1,0,2n+1} = 2^{-n-1}\pi^{-\frac{3}{2}}\beta_1^{-\frac{3}{2}}a_d^{-3}w_1\left(\frac{\beta_1\eta_{0B}^2+w_1}{2}+\frac{1}{4}\right)\times$$
$$\times\left[-\frac{\Gamma\left(\frac{\beta_1\eta_{0B}^2-w_1}{2}+\frac{3}{4}\right)}{(2n+1)!\,\Gamma\left(\frac{1}{2}-w_1\right)\Gamma\left(\frac{\beta_1\eta_{0B}^2+w_1}{2}+\frac{5}{4}\right)}\times\right.$$
$$\left.\times\frac{1}{\left[\left(\frac{\beta_1\eta_{0B}^2+w_1}{2}+\frac{1}{4}\right)\left[\Psi\left(\frac{\beta_1\eta_{0B}^2+w_1}{2}+\frac{5}{4}\right)-\Psi\left(\frac{\beta_1\eta_{0B}^2-w_1}{2}+\frac{3}{4}\right)\right]-1\right]}\right]^{\frac{1}{2}}. \tag{25}$$

Let us suppose that the dispersion $u$ of QD size arises under the phase decay process in resaturated solid solution [19, 20]. This is well described by Lifshits-Slezov formula [20, 21],

$$P(u)=\begin{cases} \dfrac{3^4 e u^2 \exp\left[-1/(1-2u/3)\right]}{2^{\frac{5}{3}}(u+3)^{\frac{7}{3}}(3/2-u)^{\frac{11}{3}}}, & u<\dfrac{3}{2}, \\ 0, & u>\dfrac{3}{2}, \end{cases} \tag{26}$$

where $u=R_0/\overline{R_0}$, $R_0$ *and* $\overline{R_0}$ are QD radius and the mean value of QD radius, respectively, and *e* is the natural logarithm base.

The light impurity absorption coefficient $K_B^{(s)}(\omega)$ in the case of longitudinal polarization $\vec{e}_{\lambda s}$, with an account for the QD size dispersion, can be represented as

$$K_B^{(s)}(\omega)=\frac{2\pi N_0}{\hbar I_0}\sum_m\sum_{n_1,n}\delta_{m,0}\int_0^{\frac{3}{2}}du\; P(u)\left|M_{f\,QD,\lambda_B}^{(s)}\right|^2\frac{1}{\hbar\omega_0(u)\beta^*\left(X-\eta_{0B}^2\right)}\times$$
$$\times\delta\left(\frac{2n+3/2+(2n_1+1)\sqrt{1+\beta^{*2}a^{*-4}u^2}}{\beta^*\left(X-\eta_{0B}^2\right)}-u\right), \tag{27}$$

where $X=\hbar\omega/E_d$ is photon energy in the effective Bohr energy units, $N_0$ is the QD concentration in dielectric matrix, $\delta(x)$ is Dirac delta-function, $\delta_{m,0}$ is Kronecker symbol, $\beta^*=\overline{R_0^*}/\left(4\sqrt{U_0^*}\right)$, and $\overline{R_0^*}=2\overline{R_0}/a_d$.

The light impurity magneto-optical absorption band edge $\hbar\omega_{thB}^{(s)}$ in the case of longitudinal polarization $\vec{e}_{\lambda s}$ is determined by the impurity level position depth, the cyclotron frequency and the QD size dispersion:

$$\hbar\omega_{thB}^{(s)} \approx E_{0,0,1}(u_0) + \left|E_{0\lambda_B}\right| . \tag{28}$$

Fig. 3 shows the photon energy cutoff value $\hbar\omega_{thB}^{(s)}$ dependence in the case of the longitudinal polarization light magneto-optical absorption for the QD-D$^{(-)}$-center complex based on InSb which is synthesized in borosilicate glass matrix, as a function of the magnetic induction *B*. As one can see from Fig. 3, this dependence is of monotonic increasing character, and the light impurity absorption band edge displacement is more than 0.03 eV, at the external magnetic field *B*=12 T.

To perform the integral in Eq. (27) it is requires to find roots of the argument of Dirac delta-function. The corresponding equation is

$$\frac{2n + 3/2 + (2n_1+1)\sqrt{1+\beta^{*2} a^{*-4} u^2}}{\beta^*\left(X - \eta_{0B}^2\right)} - u = 0 . \tag{29}$$

Roots of this equation can be found by solving the following system:

$$\begin{cases} \beta^{*2}\left[\left(X-\eta_{0B}^2\right)^2 - (2n_1+1)^2 a^{*-4}\right]u^2 - 2\beta^*(2n+3/2)\left(X-\eta_{0B}^2\right)u + \\ \qquad + (2n+3/2)^2 - (2n_1+1)^2 = 0, \\ u > \dfrac{(2n+3/2)}{\beta^*\left(X-\eta_{0B}^2\right)}, \\ X \ge X_{thB}^{(s)}. \end{cases} \tag{30}$$

where $X_{thB}^{(s)} = \hbar\omega_{thB}^{(s)} / E_d$ . It is easy to show that the solution $u_{n_1 n}$ of the system (30) is

$$u_{n_1,n} = \frac{\left(2n+\frac{3}{2}\right)\left(X-\eta_{0B}^2\right) + (2n_1+1)\sqrt{\left(X-\eta_{0B}^2\right)^2 + \left[\left(2n+\frac{3}{2}\right)^2 - (2n_1+1)^2\right]a^{*-4}}}{\beta^*\left[\left(X-\eta_{0B}^2\right)^2 - (2n_1+1)^2 a^{*-4}\right]} . \tag{31}$$

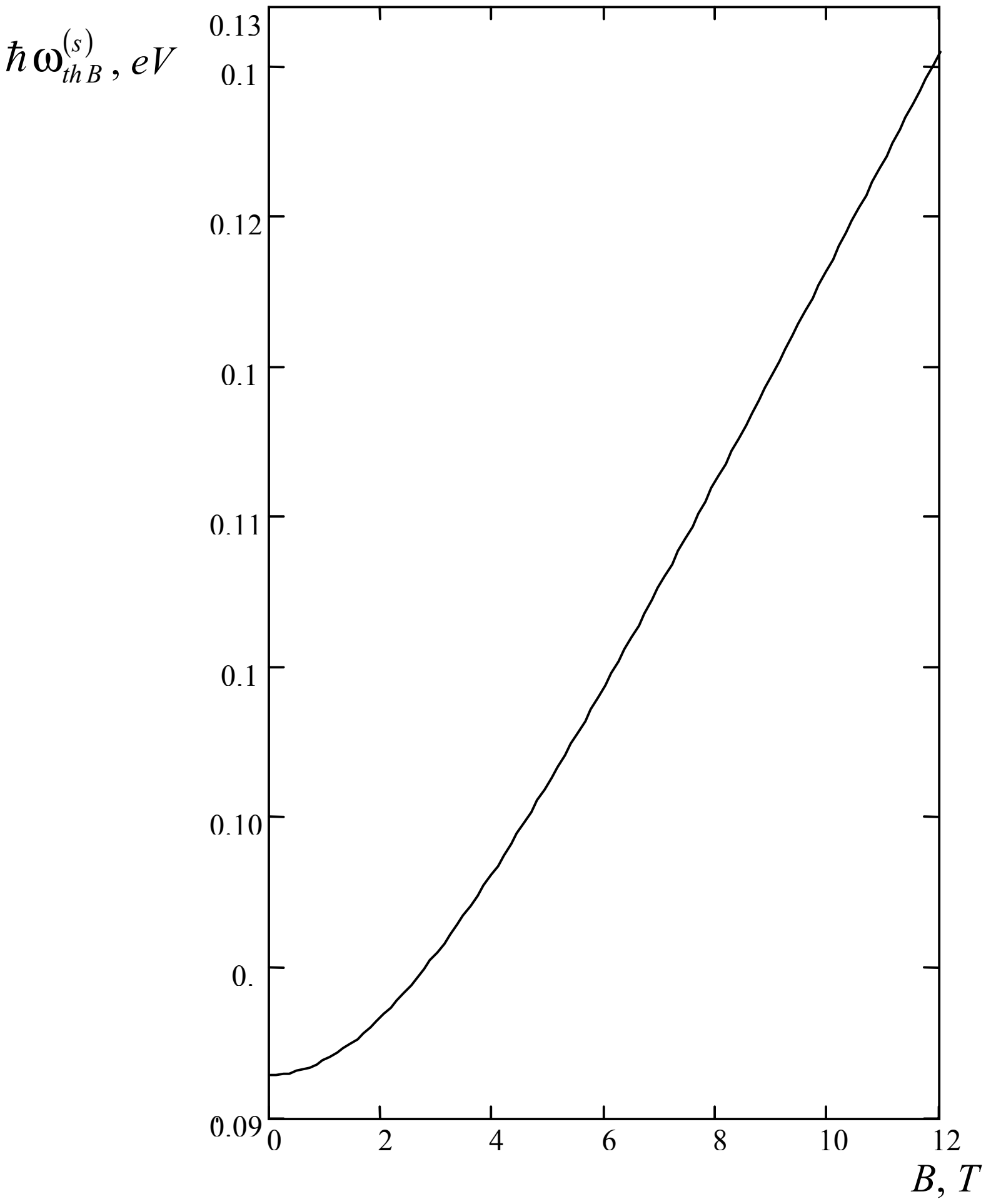

Fig. 3. The photon energy cutoff value $\hbar\omega_{thB}^{(s)}$ in the case of longitudinal polarization light magneto-optical absorption for the QD-D$^{(-)}$-center complex based on InSb, $\vec{R}_a = (0,0,0)$, $|E_i| = 5.5\times10^{-2}\ eV$, $\overline{R_0} = 71.6\ nm$, $U_0 = 0.3\ eV$, which is synthesized in borosilicate glass matrix, as a function of the magnetic induction $B$.

Using Eqs. (24)-(26) and (31) the light impurity absorption coefficient $K_B^{(s)}(\omega)$ in the case of longitudinal polarization $\vec{e}_{\lambda s}$ can be written as

$$K_B^{(s)}(\omega)=K_0\,\beta^* X\sum_{n_1=0}^{N^{(1)}}\sum_{n=0}^{N^{(2)}}u_{n_1,n}^2 P\left(u_{n_1,n}\right)\frac{(2n_1+1)(2n+2)^2\,\Gamma(2n+2)}{\Gamma^2(n+2)}\times\left(2w_{n_1,n}\right)\times$$

$$\times\left|\beta^* u_{n_1,n}\left[(2n_1+1)^2 a^{*-4}-\left(X-\eta_{0B}^2\right)^2\right]+\left(2n+\frac{3}{2}\right)\left(X-\eta_{0B}^2\right)\right|^{-1}\times$$

$$\times\frac{\Gamma\left(\delta_{n_1,n}-w_{n_1,n}+\frac{3}{4}\right)}{\left[-\Gamma\left(\frac{1}{2}-2w_{n_1,n}\right)\right]\Gamma\left(\delta_{n_1,n}+w_{n_1,n}+\frac{5}{4}\right)}\times$$

$$\times\frac{1}{\left[\left(\delta_{n_1,n}+w_{n_1,n}+\frac{1}{4}\right)\left(\Psi\left(\delta_{n_1,n}+w_{n_1,n}+\frac{5}{4}\right)-\Psi\left(\delta_{n_1,n}-w_{n_1,n}+\frac{3}{4}\right)\right)-1\right]}\times$$

$$\times\frac{\left(\delta_{n_1,n}+w_{n_1,n}+\frac{1}{4}\right)^2}{\left(\beta^{*2}u_{n_1,n}^2X^2-1\right)^2}. \qquad (32)$$

Here,

$$K_0=2^4\pi^2\lambda_0^2\alpha^* a_d^2 N_0,$$

$N^{(1)}=\left[C^{(1)}\right]$ is the integer part of

$$C^{(1)}=3\left(\beta^*\left(X-\eta_{0B}^2\right)-1\right)/\left(4\sqrt{1+9\beta^{*2}a^{*-4}/4}\right)-1/2,$$

$N^{(2)} = \left[C^{(2)}\right]$ is the integer part of

$$C^{(2)} = 3/4 \times \left(\beta^* \left(X - \eta_{0B}^2\right) - 1\right) - \left(n_1 + 1/2\right)\sqrt{1 + 9\beta^{*2} a^{*-4}/4}\,,$$

$w_{n_1,n} = \sqrt{1 + \beta^{*2} a^{*-4} u_{n_1,n}^2}/2$, and $\delta_{n_1,n} = \beta^* \eta_{0B}^2 u_{n_1,n}/2$.

Fig. 4 represents the spectral dependence of the light magneto-optical impurity absorption coefficient $K_B^{(s)}(\omega)$ for the longitudinal polarization in the case of borosilicate glass, which is pigmented by the InSb crystallites. With an increase of the magnetic field (compare curves 1 and 2 of Fig. 4), the light impurity absorption band edge shifts to the short-wave spectrum region, that is related to the corresponding dynamics of the impurity level and Landau levels. The oscillation period (see curve 2 of Fig. 4), with the quantum number $n_1$ rising or lowering by 1, is determined by the hybrid frequency,

$$\Omega = \sqrt{4\omega_0^2 + \omega_B^2}\,,$$

and equals to $\hbar\Omega$ . The distance between two nearest bands for the light impurity magneto-optical absorption longitudinal polarization spectrum is $2\hbar\omega_0$.

We now turn to the case of ***transversal*** polarization, $\vec{e}_\lambda \perp \vec{B}$, with respect to the direction of the external magnetic field. We assume that the binding energy level $E_{0\lambda_B}$ [D$^{(-)}$-center is localized at the point $\vec{R}_a = (0,0,0)$], which is lower than the QD parabolic potential well bottom, i.e., $E_{0\lambda_B} < 0$.

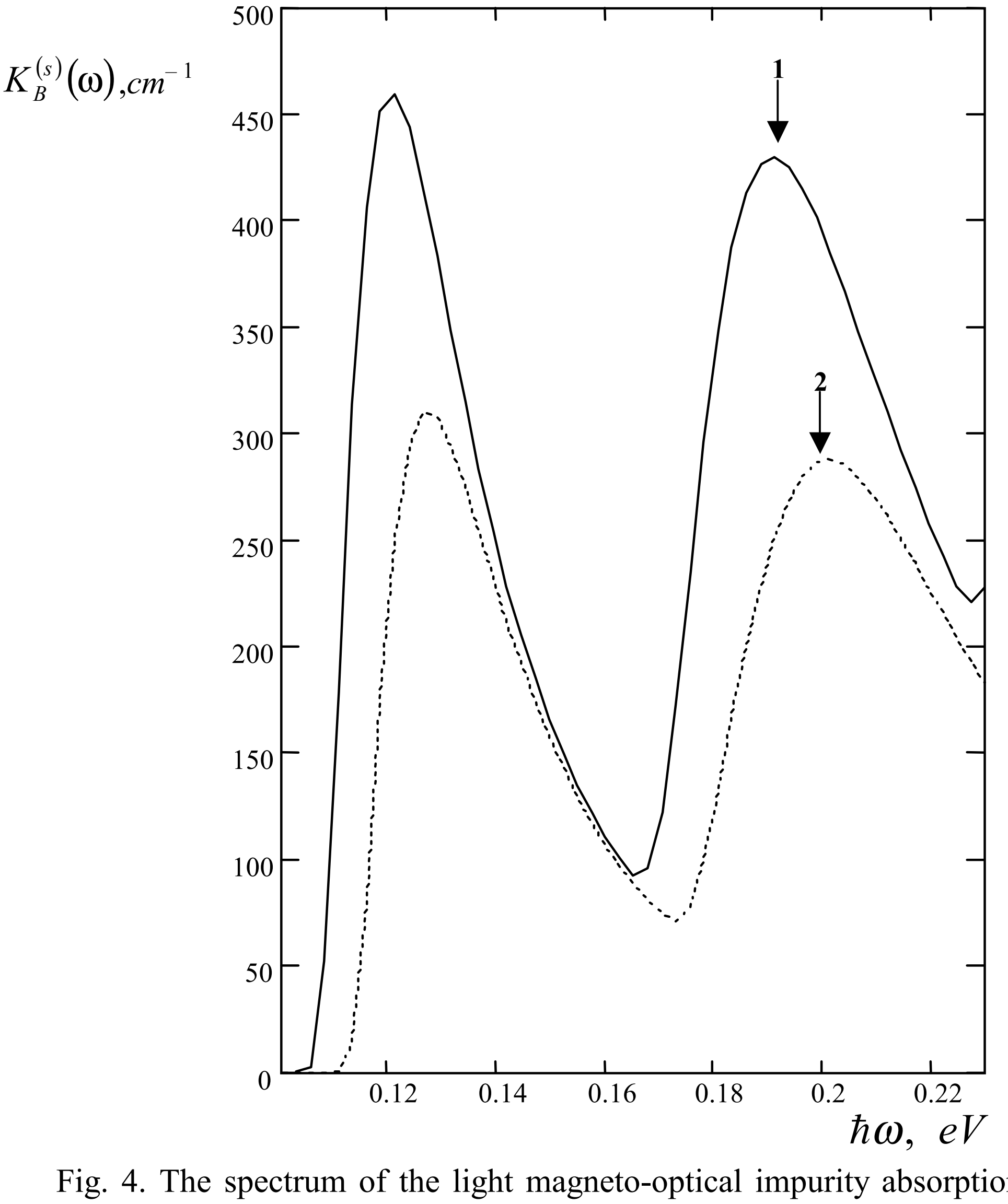

Fig. 4. The spectrum of the light magneto-optical impurity absorption coefficient $K_B^{(s)}(\omega)$ , for the longitudinal polarization in the case of borosilicate glass, which is pigmented by InSb crystallites, as a function of magnetic induction $B$; $|E_i| = 3.5\times10^{-2}$ $eV$, $\overline{R_0} = 35.8$ $nm$, $U_0 = 0.2$ $eV$, $N_0 = 10^{15}$ $sm^{-3}$. Curve 1: $B = 0$ $T$, curve 2: $B = 5$ $T$.

The effective Hamiltonian $H_{\text{int}B}^{(t)}$ of the interaction with the light wave field (wave vector $\vec{q}_t$ and the unit transversal polarization vector $\vec{e}_{\lambda t}$, with respect to the magnetic field) can be written as

$$H_{\text{int}B}^{(t)} = -i\hbar\lambda_0\sqrt{\frac{2\pi\hbar^2\alpha^*}{m^{*2}\omega}I_0}\,\exp\left(i\vec{q}_t\vec{r}\right)\left(\left(\vec{e}_{\lambda t}\nabla_{\vec{r}}\right)-\frac{i|e|B}{2\hbar}\left[\vec{e}_{\lambda t},\vec{r}\right]_z\right). \quad (33)$$

In the dipole approximation, the matrix elements $M_{fQD,\lambda_B}^{(t)}$, which determine the electron optical transitions from $D^{(-)}$-center ground state $\Psi_{\lambda_B}^{(QD)}(\rho,\varphi,z;0)$ to the states of QD discrete spectrum $\Psi_{n_1,m,n_2}(\rho,\varphi,z)$, in the case of transversal polarization $\vec{e}_{\lambda t}$, can be represented as the sum of two parts:

$$M_{fQD,\lambda_B}^{(t)} = M_1 + M_2, \quad (34)$$

where

$$M_1 = i\lambda_0\sqrt{\frac{2\pi\alpha^* I_0}{\omega}}\left(E_{n_1,m,n_2}-E_{0\lambda_B}\right)\left\langle\Psi_{n_1,m,n_2}^*(\rho,\varphi,z)\right|\left(\vec{e}_{\lambda t},\vec{r}\right)\left|\Psi_{\lambda_B}^{(QD)}(\rho,\varphi,z;0)\right\rangle, \quad (35)$$

$$M_2 = -\lambda_0\sqrt{\frac{2\pi\alpha^* I_0}{\omega}}\frac{\hbar\omega_B}{2}\left\langle\Psi_{n_1,m,n_2}^*(\rho,\varphi,z)\right|\left[\vec{e}_{\lambda t},\vec{r}\right]_z\left|\Psi_{\lambda_B}^{(QD)}(\rho,\varphi,z;0)\right\rangle. \quad (36)$$

With the use of the energy spectrum (3), the QD electron wave functions (4), and the $D^{(-)}$ -center binding state wave function, Eq. (35) can be rewritten as

$$M_1 = 2^{\frac{1}{2}}\pi^2 i\lambda_0\sqrt{\frac{\alpha^* I_0}{\omega}}\exp(\mp i\vartheta)E_d\,a_d^4\,\beta_1 w_1^{-\frac{1}{2}}(-1)^n\frac{(2n)!}{n!}C_{1B}^{QD}\,C_{n_1,\pm1,2n}\times$$
$$\times\frac{\left(m\beta_1 a^{*-2}+(2n+1/2)+(2n_1+2)w_1+\beta_1\eta_{0B}^2\right)}{\left(\frac{\beta_1\eta_{0B}^2}{2}+n+\left(n_1+\frac{1}{2}\right)w_1+\frac{1}{4}\right)\left(\frac{\beta_1\eta_{0B}^2}{2}+n+\left(n_1+\frac{3}{2}\right)w_1+\frac{1}{4}\right)}, \quad (37)$$

where $\vartheta$ is the polar angle for the transversal polarization unit vector $\vec{e}_{\lambda t}$ in the cylindrical system of reference. Calculation of the r.h.s. of Eq. (37) includes the integrals of following form:

$$\int_0^{2\pi} d\varphi \cos(\varphi-\vartheta)\exp(-im\varphi)=\begin{cases}\pi\exp(\mp i\vartheta), if\ m=\pm 1,\\ 0, if\ m\neq\pm 1\end{cases} \quad (38)$$

$$\int_{-\infty}^{+\infty} dz\ \exp\left[-\frac{z^2}{2\beta_1 a_d^2\left(1-e^{-2t}\right)}\right]H_{n_2}\left(\frac{z}{\sqrt{2\beta_1}a_d}\right)=$$

$$=\begin{cases}0, if\ n_2\neq 2n, n=0,1,2,...,\\ (-1)^n\sqrt{2\pi\beta_1}\,a_d\dfrac{(2n)!}{n!}\exp[-2nt]\sqrt{1-e^{-2t}}, if\ n_2=2n,\end{cases} \quad (39)$$

The selection rules for quantum numbers $m$ and $n_2$ follow from Eqs. (38) and (39).

The normalization constant $C_{1B}^{QD}\,C_{n_1,\pm 1,2n}$ in Eq. (37) has been found as

$$C_{1B}^{QD}C_{n_1,\pm1,2n}=\pi^{-\frac{3}{2}}\beta_1^{-\frac{3}{2}}a_d^{-3}w_1\left(\frac{\beta_1\eta_{0B}^2+w_1}{2}+\frac{1}{4}\right)\times$$

$$\times\left[-\frac{(n_1+1)}{2^{2n+1}(2n)!}\frac{\Gamma\left(\frac{\beta_1\eta_{0B}^2-w_1}{2}+\frac{3}{4}\right)}{\Gamma\left(\frac{1}{2}-w_1\right)\Gamma\left(\frac{\beta_1\eta_{0B}^2+w_1}{2}+\frac{5}{4}\right)}\times\right.$$

$$\left.\times\frac{1}{\left[\left(\frac{\beta_1\eta_{0B}^2+w_1}{2}+\frac{1}{4}\right)\left[\Psi\left(\frac{\beta_1\eta_{0B}^2+w_1}{2}+\frac{5}{4}\right)-\Psi\left(\frac{\beta_1\eta_{0B}^2-w_1}{2}+\frac{3}{4}\right)\right]-1\right]}\right]^{\frac{1}{2}}. \quad (40)$$

Using Eqs. (4) and (20), the Eq. (36) defining $M_2$ entering the matrix elements (34) can be represented as

$$M_2 = 2^{\frac{1}{2}}\pi^2 i\lambda_0\sqrt{\frac{\alpha^* I_0}{\omega}}\exp(\mp i\vartheta)E_d\, a_d^4\, m a^{*-2}\,\beta_1^2 w_1^{-\frac{1}{2}}(-1)^n\frac{(2n)!}{n!}C_{1B}^{QD}C_{n_1,\pm1,2n}\times$$
$$\times\left(\frac{\beta_1\eta_{0B}^2}{2}+n+\left(n_1+\frac{1}{2}\right)w_1+\frac{1}{4}\right)^{-1}\left(\frac{\beta_1\eta_{0B}^2}{2}+n+\left(n_1+\frac{3}{2}\right)w_1+\frac{1}{4}\right)^{-1}, \quad (41)$$

In calculating Eq. (41) we used the integral

$$\int_0^{2\pi} d\varphi\,\sin(\varphi-\vartheta)\exp(-im\varphi)=\begin{cases}\mp\pi i\exp(\mp i\vartheta),\ if\ m=\pm1,\\ 0,\ if\ m\neq\pm1.\end{cases} \quad (42)$$

Using Eqs. (34), (37) and (41), for the considered optical transitions matrix elements $M_{fQD,\lambda_B}^{(t)}$ we obtain

$$M_{fQD,\lambda_B}^{(t)} = 2^{\frac{1}{2}}\pi^2 i\lambda_0\sqrt{\frac{\alpha^* I_0}{\omega}}\exp(\mp i\vartheta)E_d\, a_d^4\,\beta_1 w_1^{-\frac{1}{2}}(-1)^n\frac{(2n)!}{n!}C_{1B}^{QD}C_{n_1,\pm1,2n}\times$$
$$\times\frac{\left(2m\beta_1 a^{*-2}+(2n+1/2)+(2n_1+2)w_1+\beta_1\eta_{0B}^2\right)}{\left(\frac{\beta_1\eta_{0B}^2}{2}+n+\left(n_1+\frac{1}{2}\right)w_1+\frac{1}{4}\right)\left(\frac{\beta_1\eta_{0B}^2}{2}+n+\left(n_1+\frac{3}{2}\right)w_1+\frac{1}{4}\right)}. \quad (43)$$

Here, $C_{1B}^{QD}C_{n_1,\pm1,2n}$ is determined by Eq. (40) and $m=\pm1$. As one can see from Eqs. (38), (39) and (42), selection rules for the magnetic quantum number $m$ ($m=\pm1$) and quantum number $n_2$ ($n_2=2\ n,\ n=0,\ 1,\ 2,...$) are such that the optical transitions from impurity level are possible only to the QD states with $m=\pm1$ and with even values of $n_2$. The light impurity absorption coefficient expression $K_B^{(t)}(\omega)$ for the transversal polarization case can be written as

$$K_B^{(t)}(\omega)=\frac{2\pi N_0}{\hbar I_0}\sum_{n_1,n}\sum_{m=-1}^{1}\delta_{|m|,1}\int_0^{\frac{3}{2}}du\,P(u)\left|M_{fQD,\lambda_B}^{(t)}\right|^2\delta\left(E_{n_1,m,2n}+\left|E_{0\lambda_B}\right|-\hbar\omega\right). \quad (44)$$

As it follows from the condition $\hbar\omega_0(u)=\dfrac{E_d}{\beta^* u}$ and Eq. (3) the eigenvalues $E_{n_1,m,2n}$ of the Hamiltonian (2) are decreasing functions of the QD size dispersion $u$ ($0<u<3/2$). These are found to be

$$E_{n_1,m,2n}=E_{n_1,m,2n}(u)=E_d\left(ma^{*-2}+\frac{2n+\frac{1}{2}+\left(2n_1+|m|+1\right)\sqrt{1+\beta^{*2}a^{*-4}u^2}}{\beta^* u}\right), \quad (45)$$

where $m=\pm 1$.

This result allows us to represent the light impurity absorption coefficient $K_B^{(t)}(\omega)$ in the form

$$K_B^{(t)}(\omega)=\frac{2\pi N_0}{\hbar I_0}\sum_{n_1,n}\sum_{m=-1}^{1}\delta_{|m|,1}\int_0^{\frac{3}{2}}du\,P(u)\left|M_{fQD,\lambda_B}^{(t)}\right|^2\frac{1}{\hbar\omega_0(u)\beta^*\left(X-\eta_{0B}^2\right)}\times$$
$$\times\delta\left(\frac{mu}{a^{*2}\left(X-\eta_{0B}^2\right)}+\frac{2n+1/2+\left(2n_1+2\right)\sqrt{1+\beta^{*2}a^{*-4}u^2}}{\beta^*\left(X-\eta_{0B}^2\right)}-u\right). \quad (46)$$

The photon energy cutoff value $\hbar\omega_{thB}^{(t)}$ for the transversal polarization light $\vec{e}_{\lambda t}$ case can be written as

$$X_{thB}^{(t)}\approx\eta_{0B}^2+\frac{1/2+2\sqrt{1+\beta^{*2}a^{*-4}u_0^2}}{\beta^* u_0}-a^{*-2}, \quad (47)$$

where $X_{thB}^{(t)}=\hbar\omega_{thB}^{(t)}/E_d$ and $u_0=3/2$.

In Fig. 5 we plot the photon energy cutoff value $\hbar\omega_{thB}^{(t)}$ in the case of transversal polarization light magneto-optical absorption by the QD-D$^{(-)}$-center complexes (based on InSb) synthesized in borosilicate glass matrix, as a function of the magnetic induction value $B$. This dependence is of a has non-monotonic character, with a pronounced minimum. The equation determining the roots of Dirac delta-function argument (in Eq. (44)) is

$$\frac{mu}{a^{*2}\left(X-\eta_{0B}^{2}\right)}+\frac{2n+1/2+\left(2n_{1}+2\right)\sqrt{1+\beta^{*2}a^{*-4}u^{2}}}{\beta^{*}\left(X-\eta_{0B}^{2}\right)}-u=0. \quad (48)$$

For the case of transversal polarization light $\vec{e}_{\lambda t}$ absorption, $X\geq X_{thB}^{(t)}$, solution of Eq. (48) is equivalent to solution of the following system:

$$\begin{cases} \beta^{*2}\left[\left(X-\eta_{0B}^{2}-ma^{*-2}\right)^{2}-\left(2n_{1}+2\right)^{2}a^{*-4}\right]u^{2}-2\beta^{*}\left(2n+1/2\right)\times \\ \quad\times\left(X-\eta_{0B}^{2}-ma^{*-2}\right)u+\left(2n+1/2\right)^{2}-\left(2n_{1}+2\right)^{2}=0, \\ u>\dfrac{\left(2n+1/2\right)}{\beta^{*}\left(X-\eta_{0B}^{2}-ma^{*-2}\right)}, \\ X\geq X_{thB}^{(t)}, \end{cases} \quad (49)$$

where $m=\pm 1$. The solutions of this system are found as

$$u_{n_1,n,m}=\frac{\left(2n+\frac{1}{2}\right)\left(X-\eta_{0B}^{2}-\frac{m}{a^{*2}}\right)+\left(2n_{1}+2\right)\sqrt{\left(X-\eta_{0B}^{2}-\frac{m}{a^{*2}}\right)^{2}+\left[\left(2n+\frac{1}{2}\right)^{2}-\left(2n_{1}+2\right)^{2}\right]\frac{1}{a^{*4}}}}{\beta^{*}\left[\left(X-\eta_{0B}^{2}-\frac{m}{a^{*2}}\right)^{2}-\left(2n_{1}+2\right)^{2}\frac{1}{a^{*4}}\right]}, \quad (50)$$

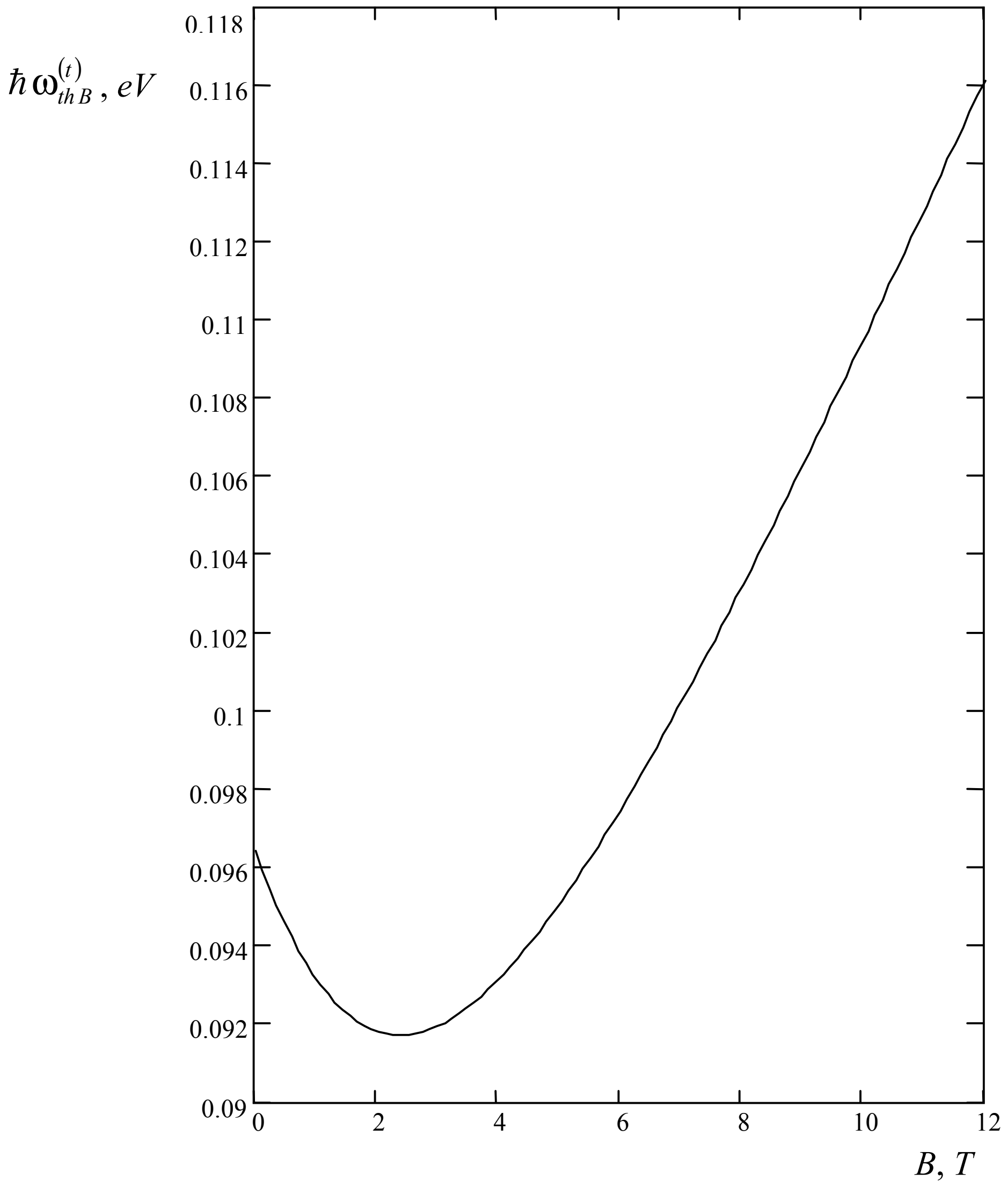

Fig. 5. The photon energy cutoff value $\hbar\omega_{thB}^{(t)}$ in the case of transversal polarization light magneto-optical absorption by the QD-D$^{(-)}$-center complexes (based on InSb) synthesized in borosilicate glass matrix, as a function of magnetic induction value $B$; $\vec{R}_a = (0,0,0)$, $|E_i| = 5.5\times10^{-2}$ $eV$, $\overline{R_0} = 71.6$ $nm$, $U_0 = 0.3$ $eV$.

$$u^{(2)}_{n_1,n,m}=\frac{\left(2n+\frac{1}{2}\right)\left(X-\eta_{0B}^2-\frac{m}{a^{*2}}\right)\mp(2n_1+2)\sqrt{\left(X-\eta_{0B}^2-\frac{m}{a^{*2}}\right)^2+\left[\left(2n+\frac{1}{2}\right)^2-(2n_1+2)^2\right]\frac{1}{a^{*4}}}}{\beta^*\left[\left(X-\eta_{0B}^2-\frac{m}{a^{*2}}\right)^2-(2n_1+2)^2\frac{1}{a^{*4}}\right]},$$

(51)

where $m=\pm 1$. It is easy to show that only the solution (50) is appropriate and, hence, it satisfies Eq. (48).

Finally, combining Eqs. (26), (43) and (50), we represent the light impurity absorption coefficient $K_B^{(t)}(\omega)$ in the case of the transversal light polarization $\vec{e}_{\lambda t}$ as

$$K^{(t)}(\omega)=K_{01}\beta^* X^{-1}\left[\left(X-a^{*-2}\right)^2\sum_{n_1=0}^{P_1^0}\sum_{n=0}^{P_2^0}\frac{\Gamma(2n+1)(n_1+1)(2n_1+2)^2}{2^{2n}\Gamma^2(n+1)}u^2_{n_1,n,-1}P\left(u_{n_1,n,-1}\right)\times\right.$$

$$\times\frac{\left(2\gamma_{n_1,n,-1}+\frac{3}{2}\right)^2\Gamma\left(\gamma_{n_1,n,-1}+\frac{1}{4}\right)\Gamma^2\left(\gamma_{n_1,n,-1}+\frac{3}{4}+n\right)}{\Gamma\left(\gamma_{n_1,n,-1}+\frac{7}{4}\right)\left[\left(\gamma_{n_1,n,-1}+\frac{3}{4}\right)\left(\Psi\left(\gamma_{n_1,n,-1}+\frac{7}{4}\right)-\Psi\left(\gamma_{n_1,n,-1}+\frac{1}{4}\right)\right)-1\right]}\times$$

$$\times\left|\beta^* u_{n_1,n,-1}\left[(2n_1+2)^2a^{*-4}-\left(X-\eta_0^2+a^{*-2}\right)^2\right]+\left(2n+\frac{1}{2}\right)\left(X-\eta_0^2+a^{*-2}\right)\right|^{-1}\times$$

$$\times\left[\sum_{k=0}^{n_1}\frac{(-1)^k C_{n_1}^k\,2^{k+1}\Gamma(k+2)}{\Gamma\left(\gamma_{n_1,n,-1}+\frac{11}{4}+n+k\right)}\frac{\left[\beta^*\left(X-\eta_0^2+a^{*-2}\right)u_{n_1,n,-1}-\left(2n+\frac{1}{2}\right)\right]^{k+\frac{3}{2}}}{\left[\beta^*\left(X-\eta_0^2+a^{*-2}\right)u_{n_1,n,-1}+2(n_1-n)+\frac{3}{2}\right]^{k+2}}\times\right.$$

$$\times F\left(\gamma_{n_1,n,-1}+\frac{3}{4}+n,k+2;\gamma_{n_1,n,-1}+\frac{11}{4}+n+k;1-\frac{2(2n_1+2)}{\beta^*\left(X-\eta_0^2+a^{*-2}\right)u_{n_1,n,-1}+2(n_1-n)+\frac{3}{2}}\right)\Bigg]^2+$$

$$+\left(X+a^{*-2}\right)^2\sum_{n_1=0}^{P_1^0}\sum_{n=0}^{P_2^0}\frac{\Gamma(2n+1)(n_1+1)(2n_1+2)^2}{2^{2n}\Gamma^2(n+1)}u_{n_1,n,+1}^2P\left(u_{n_1,n,+1}\right)\times$$

$$\times\frac{\left(2\gamma_{n_1,n,+1}+\frac{3}{2}\right)^2\Gamma\left(\gamma_{n_1,n,+1}+\frac{1}{4}\right)\Gamma^2\left(\gamma_{n_1,n,+1}+\frac{3}{4}+n\right)}{\Gamma\left(\gamma_{n_1,n,+1}+\frac{7}{4}\right)\left[\left(\gamma_{n_1,n,+1}+\frac{3}{4}\right)\left(\Psi\left(\gamma_{n_1,n,+1}+\frac{7}{4}\right)-\Psi\left(\gamma_{n_1,n,+1}+\frac{1}{4}\right)\right)-1\right]}\times$$

$$\times\left|\beta^*u_{n_1,n,+1}\left[(2n_1+2)^2a^{*-4}-\left(X-\eta_0^2-a^{*-2}\right)^2\right]+\left(2n+\frac{1}{2}\right)\left(X-\eta_0^2-a^{*-2}\right)\right|^{-1}\times$$

$$\times\left[\sum_{k=0}^{n_1}\frac{(-1)^kC_{n_1}^k2^{k+1}\Gamma(k+2)}{\Gamma\left(\gamma_{n_1,n,+1}+\frac{11}{4}+n+k\right)}\frac{\left[\beta^*\left(X-\eta_0^2-a^{*-2}\right)u_{n_1,n,+1}-\left(2n+\frac{1}{2}\right)\right]^{k+\frac{3}{2}}}{\left[\beta^*\left(X-\eta_0^2-a^{*-2}\right)u_{n_1,n,+1}+2(n_1-n)+\frac{3}{2}\right]^{k+2}}\times\right.$$

$$\left.\times F\left(\gamma_{n_1,n,+1}+\frac{3}{4}+n,k+2;\gamma_{n_1,n,+1}+\frac{11}{4}+n+k;1-\frac{2(2n_1+2)}{\beta^*\left(X-\eta_0^2-a^{*-2}\right)u_{n_1,n,+1}+2(n_1-n)+\frac{3}{2}}\right)\right]^2\Bigg],$$

(52)

where $K_0=2^4\pi^2\lambda_0^2\alpha^*a_d^2N_0$, $P_1^0=\left[C_0^{(3)}\right]$ is the integer part of

$$C_0^{(3)}=\left(3\beta^*\left(X-\eta_0^2+a^{*-2}\right)-1\right)/\left(4\sqrt{1+9\beta^{*2}a^{*-4}/4}\right)-1,$$

$P_2^0=\left[C_0^{(4)}\right]$ is the integer part of

$$C_0^{(4)}=1/4\times\left(3\beta^*\left(X-\eta_0^2+a^{*-2}\right)-1\right)-(n_1+1)\sqrt{1+9\beta^{*2}a^{*-4}/4},$$

$u_{n_1,n,\pm1}$ are determined by the formula (50), in which coefficient $\eta_{0B}^2$ should be replaced by $\eta_0^2$, and $\gamma_{n_1,n,\pm1}=\beta^*\eta_0^2u_{n_1,n,\pm1}/2$.

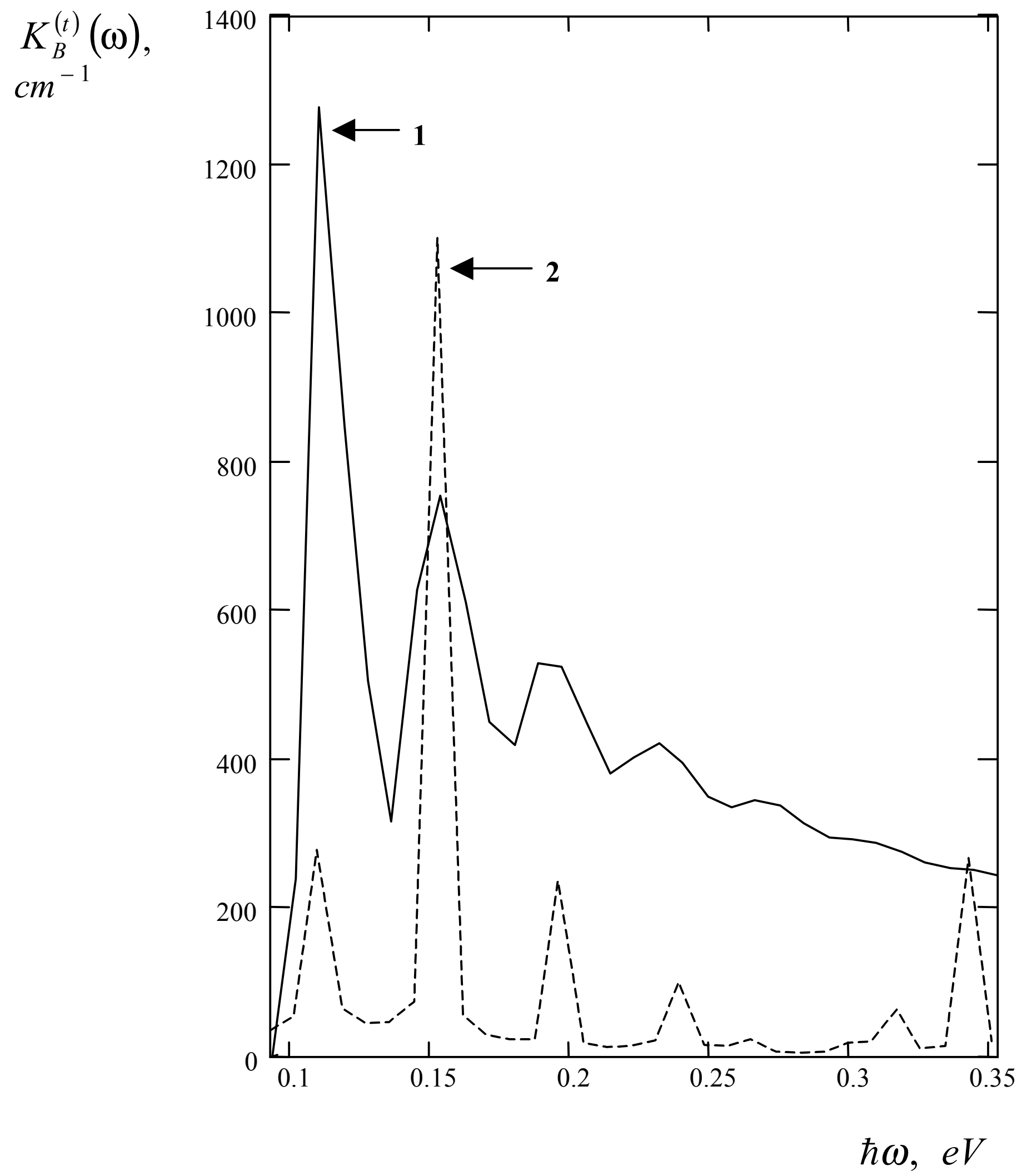

Fig. 6. The spectral dependence for the light magneto-optical impurity absorption coefficient $K_B^{(t)}(\omega)$ in the case of the transversal polarization for the borosilicate glass, which is pigmented by InSb crystallites, on magnetic induction $B$. Curve 1: $B = 0 \quad T$ and curve 2: $B = 4 \quad T$. $|E_i| = 5.5\times10^{-2} \quad eV$, $\overline{R_0} = 71.6 \quad nm$, $U_0 = 0.3 \quad eV$, $N_0 = 10^{15} \quad sm^{-3}$,

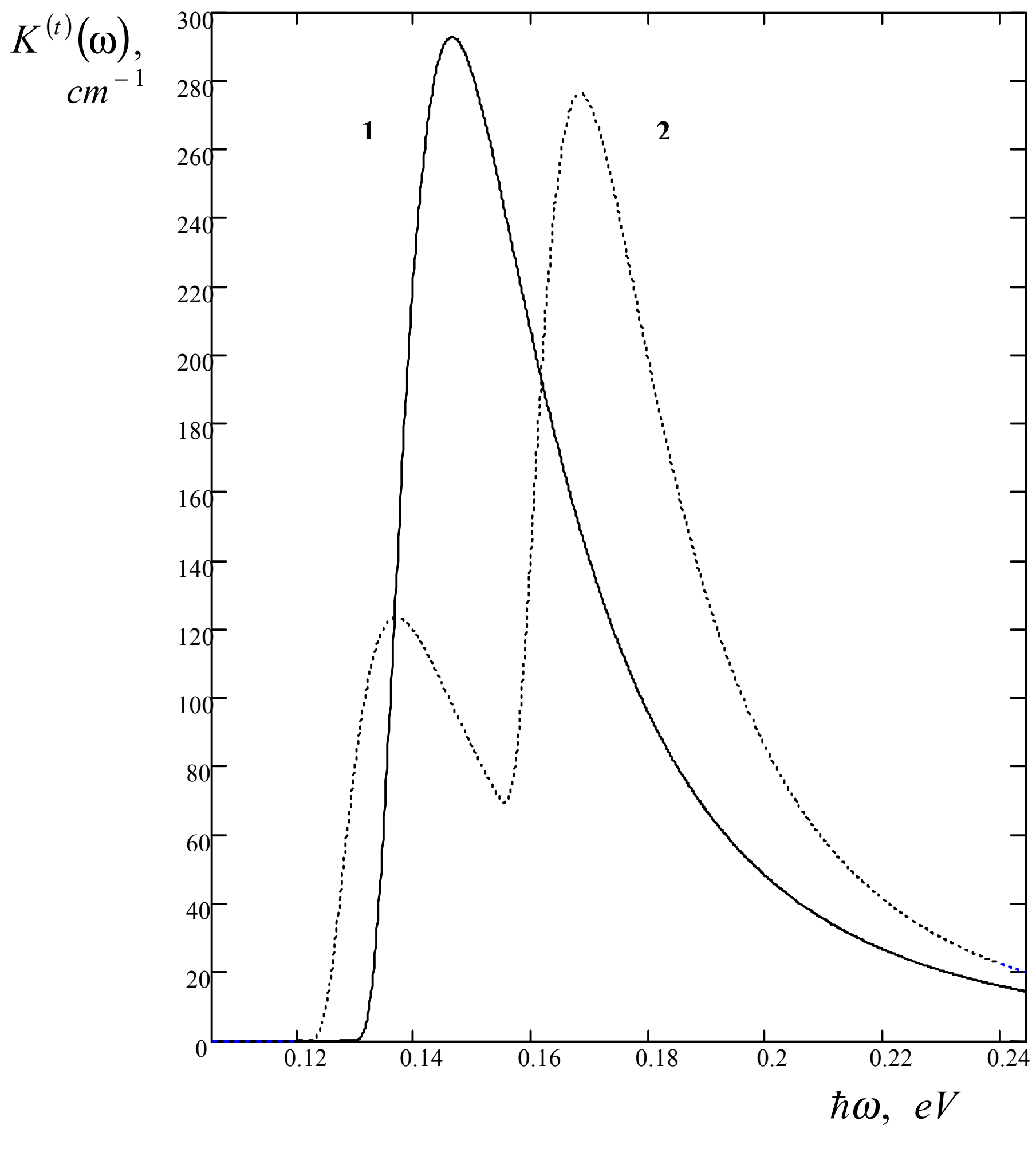

Fig. 7. The spectral dependence for the light magneto-optical impurity absorption coefficient $K_B^{(t)}(\omega)$ in the case of transversal polarization for the borosilicate glass, which is pigmented by InSb crystallites at different values of magnetic induction $B$. Curve 1: $B=0$ $T$ and curve 2: $B=3.7$ $T$. $|E_i|=5.5\times10^{-2}$ $eV$, $\overline{R_0}=35.9$ $nm$, $U_0=0.2$ $eV$, $N_0=10^{15}$ $sm^{-3}$.

Fig. 6 represents the spectral dependence of the light magneto-optical impurity absorption coefficient $K_B^{(t)}(\omega)$ in the case of transversal polarization for the borosilicate glass, which is pigmented by InSb crystallites.

Fig. 7 shows the spectral dependence of the light magneto-optical impurity absorption coefficient $K^{(t)}(\omega)$ in the case of transversal polarization when the influence of magnetic field to the QD impurity ground state is negligible. The impurity absorption band (curve 1 of Fig. 7) in external magnetic field is splitted into the Zeeman doublet (curve 2 of Fig. 7). Height of absorption peak related to the electron optical transition to the state with $m=-1$ is several times smaller than the peak related to the electron optical transition to state with $m=+1$. Such an asymmetry of the doublet is due to displacement from the spherically symmetrical potential well for the electron wave function, which corresponds to the state with $n_1=0$, $m=-1$, and $n_2=0$ [13]. The distance between the peaks in Zeeman doublet is determined by the cyclotron frequency, i.e., equals to $\hbar\omega_B$, and the distance between two nearest doublets depends on the confinement potential character frequency $\omega_0$, and equals to $2\hbar\omega_0$. For the quantum number $n_1$ changing by 1, the distance between the nearest doublets becomes $\hbar\Omega$, i.e., it is determined by the hybrid frequency.

# 4 Conclusions

In summary, using the zero-range potential model in the effective mass approximation, we found an analytically exact solution of the problem of binding states in QD-$D^{(-)}$-center complex under the influence of external magnetic field. It is found that a drastic modification of the QD electron states which is caused by the double quantization leads to a spatial anisotropy of the $D^{(-)}$-center binding energy. Namely, the dependence of the binding energy dependence on the polar radius in QD for impurity levels, which are lower than the bottom of the QD one, is similar to the corresponding dependence in QW (external magnetic field implies a stabilization of the binding states), and the impurity centers binding energy slightly decreases along the direction of the magnetic field.

We have shown that the spectral dependence of the light impurity absorption coefficient for the case of longitudinal polarization is of an oscillating character. The period of oscillations is determined by the hybrid frequency if the Landau level number is changed by 1, while with constant Landau level number the period of oscillations is determined by the oscillator character frequency.

We have also shown that the light impurity absorption coefficient spatial dependence in case of transversal polarization is characterized by the quantum-dimensional Zeeman effect with an asymmetric doublet.

It has been found that the distance between peaks in doublet is determined by the cyclotron frequency, and the distance between two nearest doublets, at the constant Landau level number, is determined by the oscillator character frequency, while when the Landau level number is changed by 1 this distance is determined by the hybrid frequency.